\documentclass[conference]{IEEEtran}
\IEEEoverridecommandlockouts
\usepackage{cite}
\usepackage{amsmath,amssymb,amsfonts}
\usepackage{algorithmic}
\usepackage{graphicx}
\usepackage{textcomp}
\usepackage[table,xcdraw]{xcolor}
\usepackage{multirow}
\usepackage{url}
\usepackage[skip=0.5\baselineskip]{caption}

\def\BibTeX{{\rm B\kern-.05em{\sc i\kern-.025em b}\kern-.08em
    T\kern-.1667em\lower.7ex\hbox{E}\kern-.125emX}}
\begin{document}

\title{A Study of Incorporating Articulatory Movement Information in Speech Enhancement
}

\author{\IEEEauthorblockN{Yu-Wen Chen$^1$, Kuo-Hsuan Hung$^1$, Shang-Yi Chuang$^1$, Jonathan Sherman$^1$, Xugang Lu$^2$, Yu Tsao$^1$}
\IEEEauthorblockA{$^1$Research Center for Information Technology Innovation, Academia Sinica, Taiwan}
\IEEEauthorblockA{$^2$National Institute of Information and Communications Technology, Japan}
}

\maketitle

\begin{abstract}

Although deep learning algorithms are widely used for improving speech enhancement (SE) performance, the performance remains limited under highly challenging conditions, such as unseen noise or noise signals having low signal-to-noise ratios (SNRs). This study provides a pilot investigation on a novel multimodal audio-articulatory-movement SE (AAMSE) model to enhance SE performance under such challenging conditions. Articulatory movement features and acoustic signals were used as inputs to waveform-mapping-based and spectral-mapping-based SE systems with three fusion strategies. In addition, an ablation study was conducted to evaluate SE performance using a limited number of articulatory movement sensors. Experimental results confirm that, by combining the modalities, the AAMSE model notably improves the SE performance in terms of speech quality and intelligibility, as compared to conventional audio-only SE baselines.

\end{abstract}

\begin{IEEEkeywords}
articulatory movement, multimodal learning, neural network, speech enhancement
\end{IEEEkeywords}

\section{Introduction}
\label{sec:intro}

Speech enhancement (SE) aims to improve speech quality and intelligibility by reducing noise components within distorted speech signals. SE is commonly used as a pre-processing method in various speech-related applications, such as automatic speech recognition (ASR) \cite{el2007evaluation, li2015robust, vincent2018audio}, speaker recognition \cite{michelsanti2017conditional}, and hearing aids \cite{levit2001noise, healy2019deep}. Recently, neural-network (NN)-based SE methods are increasingly discussed in the research field. The deep denoising autoencoder \cite{lu2013speech, xia2014wiener, shivakumar2016perception}, fully connected neural network \cite{liu2014experiments, xu2015regression, kolbk2017speech}, convolutional neural network \cite{hui2015convolutional, fu2017raw, pandey2019new}, long short-term memory model \cite{weninger2015speech, chen2015speech, sun2017multiple}, and attention-mechanism-based models \cite{kim2020transformer, koizumi2020speech, yang2020characterizing, fu2020boosting} are well-known SE methods that use NN models as the core architecture.

NN-based SE methods often only use audio signals as the input. However, the contingent weak point is that the SE performance decreases drastically when encountering unknown noise or very low signal-to-noise ratio (SNR) conditions. Hence, audio-visual multimodal SE systems were developed to address this issue \cite{hou2016audio, hou2018audio}. However, visual data have several limitations - only the external vocal tract (lips) are considered, greater storage and processing capacities are required, and unseen video conditions (capture quality/lighting, obstructions, facial angles, sudden movements, etc.) will limit performance similar to unseen audio - the same weakness it attempted to improve. Conversely, articulatory features such as broad phone class (BPC) and articulatory movements are robust to environmental changes. \cite{lu2020speech} has shown that using BPC can improve the SE performance. Also, recent studies have confirmed that articulatory movements provide useful and complementary information to acoustic signals and, hence, can be used to synthesize speech signals \cite{bocquelet2014robust, taguchi2018articulatory}. 

This study serves as a pilot investigation of the situation wherein both articulatory movements and acoustic signals are available, while acoustic signals might be distorted. Combining articulatory movements and acoustic sensors can be used to facilitate effective vocal communication in extremely noisy circumstances (sports events, factories, crowded places) with no visual data available. 

In this study, the electromagnetic midsagittal articulography (EMMA) method was used to collect articulatory movements. Note that EMMA is just one particular way to collect articulatory movements. In recent years, numerous in-mouth sensors, such as smart palate systems \cite{fabus2015preliminary, zin2021technology}, smart dental braces \cite{kutbee2017flexible}, and in-mouth monitoring \cite{ma2017wireless}, have been developed to collect articulatory features. Therefore, we are certain that in-mouth sensors will have increased practical usage in the future, and the results of this study can be applied to articulatory movements collected from various devices.

The EMMA technology captures articulatory movements by inducing current in sensors placed on articulators (tongues or lips) using an electromagnetic field. Wei \emph{et al.} \cite{wei2018study} and Chen \emph{et al.} \cite{chen2021ema2s} studied the contribution of articulators to speech. Hiroya \emph{et al.} \cite{hiroya2004estimation} used an HMM-based speech production model to estimate the articulatory movements from speech acoustics. However, to the best of our knowledge, the use of articulatory movements as an additional feature in SE systems has not been tested yet.

We test audio-articulatory-movement SE (AAMSE) models with three fusion strategies on both waveform-mapping-based and spectral-mapping-based SE systems. Experimental results showed that the proposed AAMSE models outperformed the baseline audio-only SE models and achieved higher intelligibility even at low SNR levels.

The remainder of this paper is organized as follows. Section \ref{sec:related} introduces the related works of this study. Section \ref{sec:proposed} presents the proposed articulatory movement features and AAMSE frameworks. Section \ref{sec:experiment} provides the experimental details and results. Finally, Section \ref{sec:conclusion} presents the conclusion of this study.

\section{Related works}
\label{sec:related}

The AAMSE was implemented on one waveform-mapping-based and two spectral-mapping-based SE systems. Fully convolutional neural networks (FCN) have been confirmed as an effective waveform-mapping-based SE model \cite{fu2017raw}. In this study, we integrate the articulatory movements in the time domain with this model. We also implement two spectral-mapping-based models: the time delay neural network (TDNN) \cite{peddinti2015time} and bi-directional long short-term memory networks (BLSTM). The two models both consider the temporal relation within speech signals. The TDNN is a fully connected feed-forward neural network that has been proven robust in handling temporal dependencies. The BLSTM network considers both forward and backward sequences of inputs and has feedback connections. Hence, the BLSTM can extend attention over arbitrary time intervals and is suitable to process time series data, such as speech signals and articulatory movements.

For the waveform-mapping-based systems, SE directly processes speech waveforms. For the spectral-mapping-based systems, short-time Fourier transform (STFT) and inverse STFT are applied to transform speech between waveforms and spectral features, where only the magnitude components are enhanced, while the phase components are borrowed from the original noisy speech.

\section{Proposed AAMSE}
\label{sec:proposed}

In this section, we first explain the EMMA signals used as articulatory movement data, followed by introducing the proposed AAMSE system with three fusion strategies.

\subsection{Characteristics of the articulatory movement data}

\begin{figure}[ht]
\centerline{\includegraphics[scale=0.8]{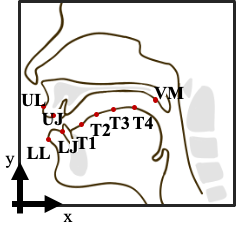}}
\caption{Positions of the EMMA sensors.}
\label{emma_fig_1}
\end{figure}

\begin{figure}[ht]
\centerline{\includegraphics[scale=0.9]{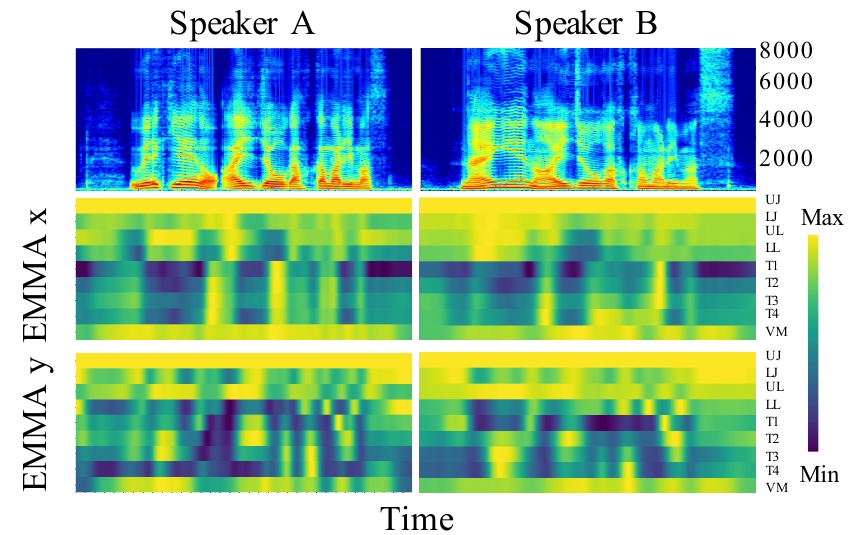}}
\caption{Visualization of the EMMA data.}
\label{emma_spec}
\end{figure}

We used the EMMA collected by NTT, Tokyo, Japan \cite{okadome2001generation} in this study. The sensor coils of the EMMA were placed on the upper lip (UL), lower lip (LL), upper jaw (UJ), lower jaw (LJ), tongue tip (T1), tongue blade (T2), tongue dorsum (T3), tongue rear (T4), and the velum (VM), as shown in Fig.~\ref{emma_fig_1}). The EMMA records the Cartesian coordinates of each sensor point at a sampling rate of 250 Hz. Fig.~\ref{emma_spec} shows the speech spectrograms and the EMMA signals of two speakers speaking the same utterance. Both the spectrograms and EMMA signals display similar patterns, indicating that these resultant signals are highly dependent on the pronunciation.

\subsection{Three fusion strategies of the AAMSE}

\begin{figure}[htbp]
\centerline{\includegraphics[scale=0.8]{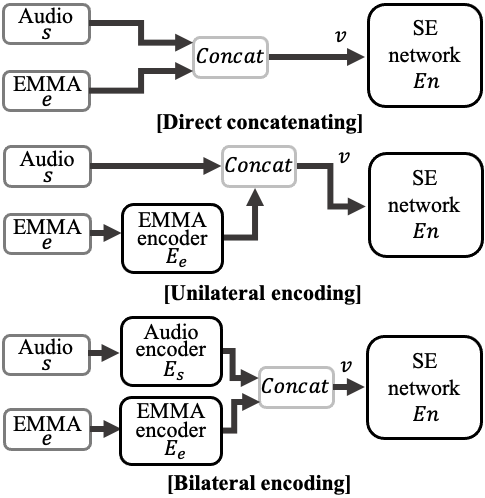}}
\caption{The three fusion strategies. The encoders and SE networks are FCN, TDNN, or BLSTM.}
\label{pro_fig_1}
\end{figure}

The aim of using SE is to convert a noisy speech signal $s$ into an enhanced speech signal $\hat{x}$ that is close to the clean speech signal $x$. We define $s$ as: $s=x+n$, where $n$ represents the noise signal.

The AAMSE is a multimodal problem. We reason that combining the physical characteristic of an audio signal and an articulatory movement, which is sound intensity and trajectory of the organs in the vocal tract, respectively, can improve the performance over the single audio modality considering they both carry speech information. We tested three fusion strategies: (1) direct concatenating, (2) unilateral encoding, and (3) bilateral encoding to integrate audio signals and articulatory movement data. Fig.~\ref{pro_fig_1} illustrates the structure of the three fusion strategies. The audio and EMMA signals are denoted by $s$ and $e$, respectively, and $v$ is the input of the SE model. The aim was to find an audio encoder $E_{s}$, an EMMA encoder $E_{e}$, and a SE network $En$ such that the enhanced signal $\hat{x}=En(v)$ was as close as possible to the clean signal $x$.

\begin{itemize}
\item Direct concatenating:
\begin{equation}
v=Concat(s,e)
\end{equation}
\item Unilateral encoding:
\begin{equation}
v=Concat(s,E_{e}(e))
\end{equation}
\item Bilateral encoding:
\begin{equation}
v=Concat(E_{s}(s),E_{e}(e))
\end{equation}
\end{itemize}

The EMMA encoder $E_{e}$, audio encoder $E_{s}$, and SE network $En$ are built by FCN, TDNN, or BLSTM. That is, we test three fusion strategies under three model structures with a total of nine combinations of the AAMSE architecture.

\section{Experiments}
\label{sec:experiment}

\subsection{Experimental setup}

The EMMA dataset comprises articulatory and speech signals from three speakers providing 354 utterances each. The two signals were recorded simultaneously at sampling rates of 250 Hz and 16 kHz for EMMA and speech, respectively. The training and testing sets included 304 and 50 utterances, respectively, from each speaker. Additionally, 100 different noise samples \cite{hu2004100} were used to prepare the noisy training data using eight different SNR levels ($\pm$ 1 dB, $\pm$ 4 dB, $\pm$ 7 dB, and $\pm$ 10 dB). Each clean utterance in the training data was contaminated with five randomly selected noises at the eight SNR levels. Similarly, each clean utterance in the testing data was corrupted with seven new noises (car noise, engine noise, pink noise, white noise, background talkers, and two types of street noises) at six different SNR levels (-8, -5, -2, 0, 2, and 5 dB).

The experimental results were evaluated using PESQ \cite{rix2001perceptual} and STOI \cite{taal2011algorithm} methods for speech quality and intelligibility, respectively. The further verified the results on a pre-trained ASR system \cite{ref_google_asr} and calculated the character correct rate (CCR) using the Levenshtein distance function \cite{levenshtein1966binary}.

\subsection{Implementation details}

The structural parameters of the waveform-mapping-based and spectral-mapping-based SE systems are listed in Table \ref{table:structures}. All waveform-mapping-based FCN \cite{fu2017raw} models were trained with L2 loss and Adam optimizer \cite{kingma2015adam} at a learning rate of 0.001. For the spectral-mapping-based models, we used STFT with a window size of 512, hop length of 128, and log1p magnitude spectrograms \cite{chuang2020lite} as the audio input feature. All spectral-mapping-based TDNN \cite{peddinti2015time} and BLSTM models were trained with L1 loss and Adam optimizer \cite{kingma2015adam} at a learning rate of 0.0001. For each SE model, we keep the same SE network structure under the audio-only condition and the audio-articulatory-movement condition with the fusion strategy of direct concatenating.

\begin{table}[!b]
\centering
\resizebox{8.8cm}{!}{
{\Large
\begin{tabular}{|c|c|c|c|}
\hline
 &
  Audio encoder &
  EMMA encoder &
  SE network \\ \hline
  
\multicolumn{4}{|c|}{FCN} \\ \hline

 \begin{tabular}[c]{@{}c@{}}Audio\\ only\end{tabular} & 
      - &
      - &
  \begin{tabular}[c]{@{}c@{}}
      Conv1d($f$:128, $k$:55)$\times$7
      \\Conv1d($f$:1, $k$:55)
  \end{tabular} \\ \hline
  \begin{tabular}[c]{@{}c@{}}Direct\\ concatenating\end{tabular}&
      - &
      - &
  \begin{tabular}[c]{@{}c@{}}
      Conv1d($f$:128, $k$:55)$\times$7
      \\ Conv1d($f$:1, $k$:55)
  \end{tabular} \\ \hline
  \begin{tabular}[c]{@{}c@{}}Unilateral\\ encoding\end{tabular} &
    - &
  \begin{tabular}[c]{@{}c@{}}
      Conv1d($f$:128, $k$:256)
      \\ Conv1d($f$:128, $k$:128)
      \\ Conv1d($f$:1, $k$:55)
  \end{tabular}&
  \begin{tabular}[c]{@{}c@{}}
      Conv1d($f$:128, $k$:55$\times$4
      \\ Conv1d($f$:1, $k$:55)
     \end{tabular} \\ \hline
  \begin{tabular}[c]{@{}c@{}}Bilateral\\ encoding\end{tabular} &
  \begin{tabular}[c]{@{}c@{}}Conv1d($f$:128, $k$:55)\\ Conv1d($f$:128, $k$:55)\\ Conv1d($f$:18, $k$:55)\end{tabular} &
  \begin{tabular}[c]{@{}c@{}}Conv1d($f$:128, $k$:128)\\ Conv1d($f$:128, $k$:128)\\ Conv1d($f$:18, $k$:64)\end{tabular} &
  \begin{tabular}[c]{@{}c@{}}Conv1d($f$:128, $k$:55)$\times$4\\ Conv1d($f$:1, $k$:55)\end{tabular} \\ \hline

\multicolumn{4}{|c|}{TDNN} \\ \hline

\begin{tabular}[c]{@{}c@{}}Audio\\ only\end{tabular} &
  - &
  - &
  \begin{tabular}[c]{@{}c@{}}TDNN(257)$\times$3   \\ Dense(771) \\ Dense(257)  \\ TDNN(257) $\times$4 \end{tabular} \\ \hline

\begin{tabular}[c]{@{}c@{}}Direct\\ concatenating\end{tabular} &
  - &
  - &
  \begin{tabular}[c]{@{}c@{}}TDNN(257)$\times$3\\ Dense(771)  \\ Dense(257)\\ TDNN(257)$\times$4\end{tabular} \\ \hline
\begin{tabular}[c]{@{}c@{}}Unilateral\\ encoding\end{tabular} &
  - &
  TDNN(18)$\times$2 &
  \begin{tabular}[c]{@{}c@{}}TDNN(257)$\times$2 \\ Dense(771)\\ Dense(257)  \\ TDNN(257)$\times$4\end{tabular} \\ \hline
\begin{tabular}[c]{@{}c@{}}Bilateral\\ encoding\end{tabular} &
  TDNN(257) &
  TDNN(18)$\times$2 &
  \begin{tabular}[c]{@{}c@{}}TDNN(257)$\times$2\\ Dense(771)\\ Dense(257) \\ TDNN(257)$\times$3 \end{tabular} \\ \hline

\multicolumn{4}{|c|}{BLSTM} \\ \hline
\begin{tabular}[c]{@{}c@{}}Audio\\ only\end{tabular} &
  - &
  - &
  \begin{tabular}[c]{@{}c@{}}BLSTM(500)$\times$3\\  Dense(257)\end{tabular} \\ \hline
\begin{tabular}[c]{@{}c@{}}Direct\\ concatenating\end{tabular} &
  - &
  - &
\begin{tabular}[c]{@{}c@{}}BLSTM(500)$\times$3\\ Dense(257)\end{tabular} \\ \hline
\begin{tabular}[c]{@{}c@{}}Unilateral\\ encoding\end{tabular} &
  - &
  \begin{tabular}[c]{@{}c@{}}BLSTM(36)$\times$3   \\ Dense(36)$\times$2\end{tabular} &
  \begin{tabular}[c]{@{}c@{}}BLSTM(514)$\times$2   \\ BLSTM(257)   \\ Dense(257)   \end{tabular} \\ \hline
\begin{tabular}[c]{@{}c@{}}Bilateral\\ encoding\end{tabular} &
  \begin{tabular}[c]{@{}c@{}}BLSTM(257)\\ Linear(257)\end{tabular} &
\begin{tabular}[c]{@{}c@{}}BLSTM(18)$\times$4   \\ Dense(18)  \end{tabular} &
\begin{tabular}[c]{@{}c@{}}BLSTM(514)$\times$2\\ BLSTM(257)\\ Dense(257)\end{tabular} \\ \hline
\end{tabular}
}
}
\caption{Waveform-mapping-based and spectral-mapping-based SE system structures. In waveform-mapping-based FCN \cite{fu2017raw}, $f$ and $k$ are the number of the output filters and kernel size, respectively. In spectral-mapping-based TDNN \cite{peddinti2015time} and BLSTM, the numbers in the brackets represent the output size.}
\label{table:structures}
\end{table}

\subsection{Experimental results}

\begin{figure}[htbp]
\centerline{\includegraphics[scale=0.85]{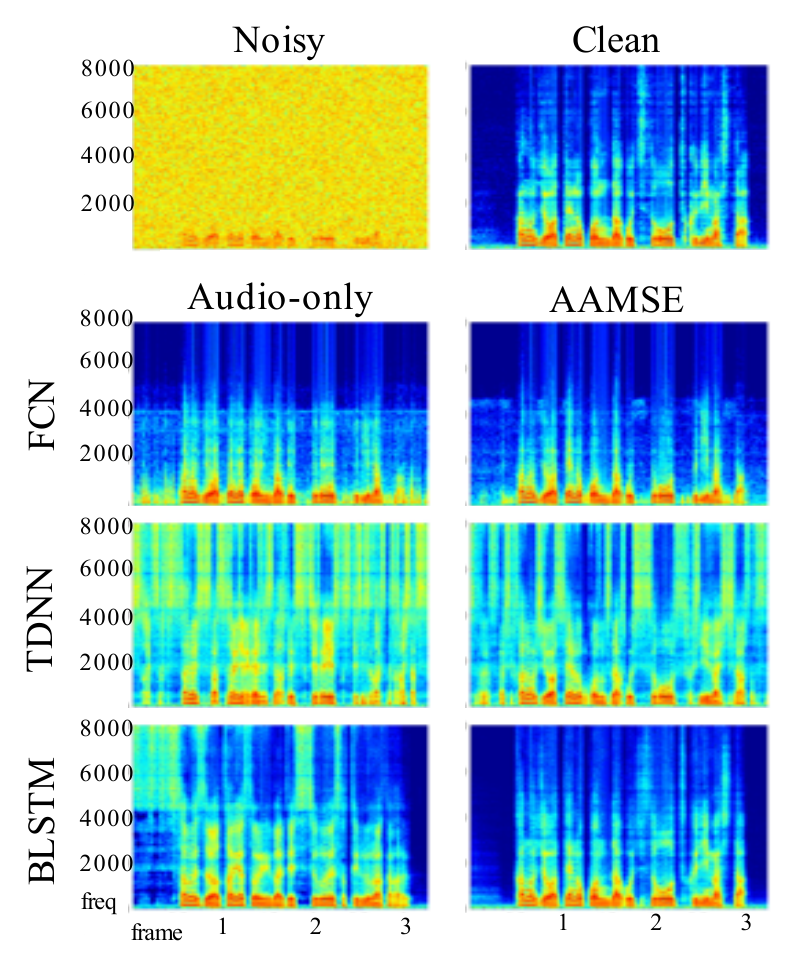}}
\caption{Spectrograms of the audio signals.}
\label{spec}
\end{figure}

The spectrograms of the enhanced audio signals in Fig.~\ref{spec} show distortion reduction in all the models. Also, as observed in the silent region, the AAMSE models show improved results than the audio-only SE baselines.

\begin{table}[b]
\centering
\resizebox{6.0cm}{!}{
{\Large
\begin{tabular}{|c|c|c|c|c|}
\hline
\multirow{2}{*}{} & \multirow{2}{*}{Noisy} & \multicolumn{3}{c|}{Audio-only}            \\ \cline{3-5} 
                  &                        & FCN            & TDNN     & BLSTM          \\ \hline
PESQ              & 1.530                  & 2.311          & 2.064    & \textbf{2.329} \\ \hline
STOI              & 0.686                  & \textbf{0.814} & 0.738    & 0.801          \\ \hline
\end{tabular}
}}
\caption{PESQ and STOI of different audio-only SE models.}
\label{table:noisy_ao}
\end{table}

The PESQ and STOI of the original noisy speech and audio-only baselines are listed in Table \ref{table:noisy_ao}. All waveform-mapping-based and spectral-mapping-based audio-only SE systems yielded higher scores than the original noisy speech. Tables \ref{table:pesq} and \ref{table:stoi} present the average scores (white part) and improvement (gray part, compared to audio-only models) in the PESQ and STOI metrics. All AAMSE models achieved higher scores than the audio-only SE models, except the FCN with unilateral encoding owing to the information loss due to channel reduction. Because SE is an audio-dominant task, we set the EMMA channel number to less than or equal to the number of audio channels. The unilateral EMMA encoder encoded EMMA signals from 18 channels to a single channel of the same size as the audio signals. Conversely, the bilateral EMMA encoder encoded EMMA signals in 18 channels without channel reduction.

\begin{table}[b]
\centering
\resizebox{8.8cm}{!}{
{\Large
\begin{tabular}{|c|c|l|c|l|c|l|c|l|}
\hline
 &
  \multicolumn{2}{c|}{\begin{tabular}[c]{@{}c@{}}Audio\\ only\end{tabular}} &
  \multicolumn{2}{c|}{\begin{tabular}[c]{@{}c@{}}Direct\\ concatenating\end{tabular}} &
  \multicolumn{2}{c|}{\begin{tabular}[c]{@{}c@{}}Unilateral \\ encoding\end{tabular}} &
  \multicolumn{2}{c|}{\begin{tabular}[c]{@{}c@{}}Bilateral\\ encoding\end{tabular}} \\ \hline
FCN   & \multicolumn{2}{c|}{2.311} & 2.653 & \cellcolor[HTML]{EFEFEF}+0.342 & 2.251 & \cellcolor[HTML]{EFEFEF}-0.060 & 2.575 & \cellcolor[HTML]{EFEFEF}+0.264 \\ \hline
TDNN  & \multicolumn{2}{c|}{2.064} & 2.402 & \cellcolor[HTML]{EFEFEF}+0.338 & 2.434 & \cellcolor[HTML]{EFEFEF}+0.370 & 2.390 & \cellcolor[HTML]{EFEFEF}+0.326 \\ \hline
BLSTM & \multicolumn{2}{c|}{2.329} & 2.793 & \cellcolor[HTML]{EFEFEF}+0.464 & \textbf{2.839} & \cellcolor[HTML]{EFEFEF}\textbf{+0.510} & 2.470 & \cellcolor[HTML]{EFEFEF}+0.141 \\ \hline
\end{tabular}
}
}

\caption{PESQ of different SE models (noisy=1.530).}
\label{table:pesq}
\end{table}

\begin{table}[!b]
\centering
\resizebox{8.8cm}{!}{
{\Large
\begin{tabular}{|c|c|l|c|c|c|c|c|c|}
\hline
 &
  \multicolumn{2}{c|}{\begin{tabular}[c]{@{}c@{}}Audio\\ only\end{tabular}} &
  \multicolumn{2}{c|}{\begin{tabular}[c]{@{}c@{}}Direct\\ concatenating\end{tabular}} &
  \multicolumn{2}{c|}{\begin{tabular}[c]{@{}c@{}}Unilateral \\ encoding\end{tabular}} &
  \multicolumn{2}{c|}{\begin{tabular}[c]{@{}c@{}}Bilateral\\ encoding\end{tabular}} \\ \hline
FCN   & \multicolumn{2}{c|}{0.814} & 0.881 & \cellcolor[HTML]{EFEFEF}+0.067 & 0.796 & \cellcolor[HTML]{EFEFEF}-0.018 & 0.862 & \cellcolor[HTML]{EFEFEF}+0.048 \\ \hline
TDNN  & \multicolumn{2}{c|}{0.738} & 0.816 & \cellcolor[HTML]{EFEFEF}+0.078 & 0.827 & \cellcolor[HTML]{EFEFEF}+0.089 & 0.820 & \cellcolor[HTML]{EFEFEF}+0.082 \\ \hline
BLSTM & \multicolumn{2}{c|}{0.801} & 0.885 & \cellcolor[HTML]{EFEFEF}+0.084 & \textbf{0.891} & \cellcolor[HTML]{EFEFEF}\textbf{+0.090} & 0.825 & \cellcolor[HTML]{EFEFEF}+0.024 \\ \hline
\end{tabular}
}
}
\caption{STOI of different SE models (noisy=0.686).}
\label{table:stoi}
\end{table}

\begin{figure}[ht]
\centerline{\includegraphics[scale=0.9]{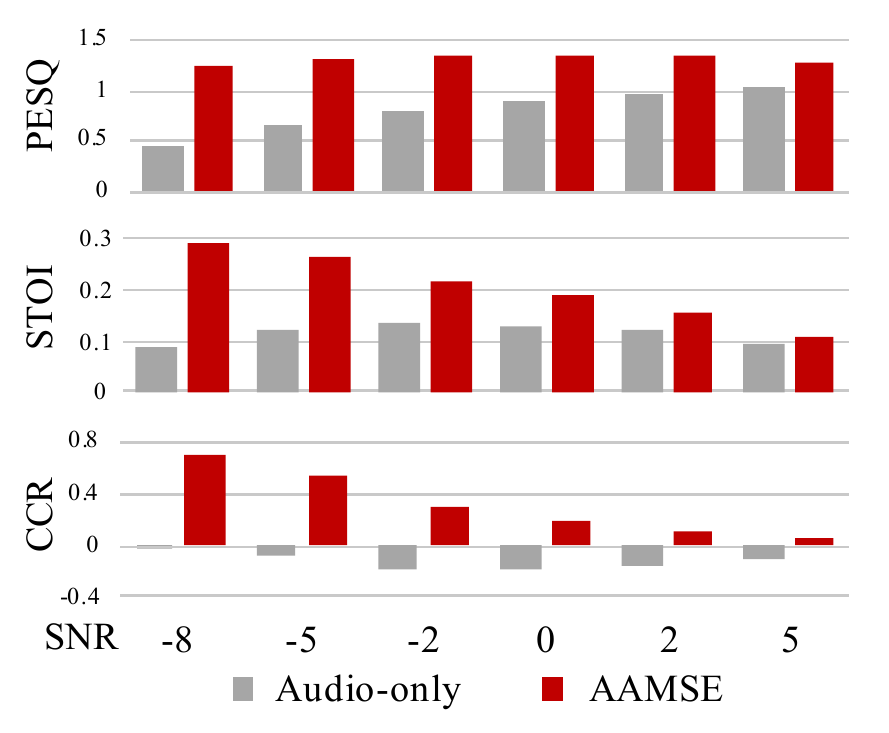}}
\caption{The SE improvement of the best audio-only SE model (BLSTM) and the best AAMSE model (BLSTM with unilateral encoding) at different SNRs.}
\label{fig_table5}
\end{figure}

\begin{figure}[!ht]
\centerline{\includegraphics[scale=0.9]{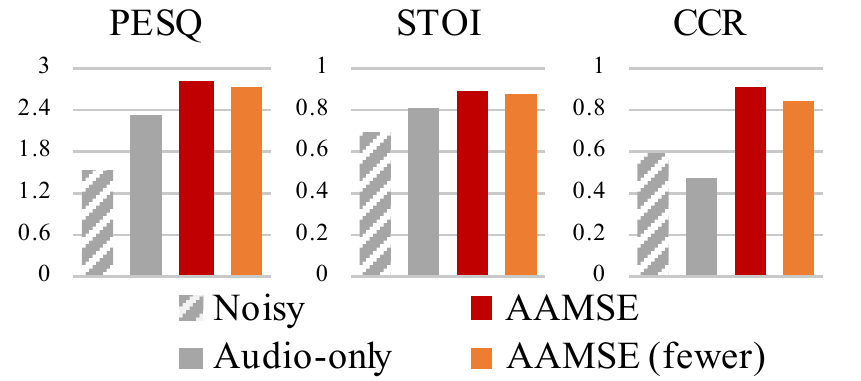}}
\caption{Average scores of different SE systems.}
\label{fig_table6}
\end{figure}

Fig.~\ref{fig_table5} shows the SE improvement ability of the best audio-only SE model (BLSTM) and best AAMSE model (BLSTM with unilateral encoding) compared to that of the original noisy signals at different SNR levels. The performance of both models improved in terms of PESQ and STOI, whereas the AAMSE model outperformed the audio-only SE model. The CCR of the audio-only SE model decreased, as reported in \cite{donahue2018exploring}, while that of the AAMSE model increased, indicating that the articulatory movement features tend to provide more information regarding intelligibility. We tested the performance of the BLSTM with unilateral encoding with four less invasive sensors (i.e., UL, LL, LJ, and T1). The experimental results, as observed in Fig.~\ref{fig_table6}, showed that the AAMSE (fewer) model achieves better performance than the audio-only SE model, indicating that a lesser combination of articulatory movement features may be sufficient for SE tasks.

\section{Conclusion}
\label{sec:conclusion}

In this study, we proposed AAMSE to enhance SE performance by incorporating articulatory movement information with acoustic signals. The experimental results showed that articulatory movements effectively improved SE performance, especially at low SNR levels. The contributions of this study are twofold: First, we confirmed the effectiveness of incorporating articulatory movements into SE systems. Second, we verified that the extra articulatory features can provide useful information for SE tasks even with only four sensors. The results of this study are promising and serve as a useful guide for designing articulatory movement data collection devices. Furthermore, we believe that the proposed AAMSE can be realized in challenging situations where speech signals are highly distorted.

\section{Acknowledgement}

The authors would like to thank the NTT Communication Science Laboratories for permitting us to use their articulatory data.

\bibliographystyle{IEEEtran}
\bibliography{refs}

\end{document}